\documentclass[twocolumn,prl,showpacs,superscriptaddress,preprintnumbers,amsmath,amssymb,floatfix]{revtex4}
\usepackage{graphicx}
\usepackage{dcolumn}
\usepackage{bm}
\usepackage{color}

\usepackage{floatrow}

\begin{document}

\title{Long-range interactions in the effective low energy Hamiltonian of Sr$_2$IrO$_4$:\\a core level resonant inelastic x-ray scattering study}
\author{S.~Agrestini}
 \affiliation{Max Planck Institute for Chemical Physics of Solids, N\"othnitzerstr. 40, 01187 Dresden, Germany}
\author{C.-Y.~Kuo}
 \affiliation{Max Planck Institute for Chemical Physics of Solids, N\"othnitzerstr. 40, 01187 Dresden, Germany}
\author{M.~Moretti~Sala}
 \affiliation{ESRF-The European Synchrotron, 71 Avenue des Martyrs, 38000 Grenoble, France}
\author{Z.~Hu}
 \affiliation{Max Planck Institute for Chemical Physics of Solids, N\"othnitzerstr. 40, 01187 Dresden, Germany}
\author{D.~Kasinathan}
 \affiliation{Max Planck Institute for Chemical Physics of Solids, N\"othnitzerstr. 40, 01187 Dresden, Germany}
\author{K.-T.~Ko}
 \affiliation{Max Planck Institute for Chemical Physics of Solids, N\"othnitzerstr. 40, 01187 Dresden, Germany}
 \affiliation{MPPC-CPM $\&$ Department of Physics, Pohang University of Science and Technology, Pohang 790-784, Korea}
\author{P.~Glatzel}
 \affiliation{ESRF-The European Synchrotron, 71 Avenue des Martyrs, 38000 Grenoble, France}
\author{M.~Rossi}
 \affiliation{ESRF-The European Synchrotron, 71 Avenue des Martyrs, 38000 Grenoble, France}
\author{J.-D.~Cafun}
 \affiliation{ESRF-The European Synchrotron, 71 Avenue des Martyrs, 38000 Grenoble, France}
\author{K.~O.~Kvashnina}
 \affiliation{ESRF-The European Synchrotron, 71 Avenue des Martyrs, 38000 Grenoble, France}
 \affiliation{Helmholtz-Zentrum Dresden-Rossendorf (HZDR), Institute of Resource Ecology – P.O. Box 510119, 01314 Dresden, Germany}
\author{A.~Matsumoto}
 \affiliation{Department of Physics and Department of Advanced Materials, University of Tokyo, 7-3-1 Hongo, Tokyo 113-0033, Japan}
\author{T.~Takayama}
 \affiliation{Department of Physics and Department of Advanced Materials, University of Tokyo, 7-3-1 Hongo, Tokyo 113-0033, Japan}
 \affiliation{Max Planck Institute for Solid State Research, Heisenbergstrasse 1, 70569 Stuttgart, Germany}
\author{H.~Takagi}
 \affiliation{Department of Physics and Department of Advanced Materials, University of Tokyo, 7-3-1 Hongo, Tokyo 113-0033, Japan}
 \affiliation{Max Planck Institute for Solid State Research, Heisenbergstrasse 1, 70569 Stuttgart, Germany}
 \affiliation{Institute for Functional Matter and Quantum Technologies, University of Stuttgart, Pfaffenwaldring 57, 70569 Stuttgart, Germany}
\author{L.~H.~Tjeng}
 \affiliation{Max Planck Institute for Chemical Physics of Solids, N\"othnitzerstr. 40, 01187 Dresden, Germany}
\author{M.~W.~Haverkort}
 \affiliation{Max Planck Institute for Chemical Physics of Solids, N\"othnitzerstr. 40, 01187 Dresden, Germany}
 \affiliation{Institute for theoretical physics, Heidelberg University, Philosophenweg 19, 69120 Heidelberg, Germany}

\date{\today}
\begin{abstract}

We have investigated the electronic structure of Sr$_2$IrO$_4$ using core level resonant inelastic x-ray scattering. The experimental spectra can be well reproduced using \textit{ab initio} density functional theory based multiplet ligand field theory calculations, thereby validating these calculations. We found that the low-energy, effective Ir $\mathrm{t}_{\mathrm{2g}}$ orbitals are practically degenerate in energy. We uncovered that covalency in Sr$_2$IrO$_4$, and generally in iridates, is very large with substantial oxygen ligand hole character in the Ir $\mathrm{t}_{\mathrm{2g}}$ Wannier orbitals. This has far reaching consequences, as not only the onsite crystal-field energies are determined by the long range crystal-structure, but, more significantly, magnetic exchange interactions will have long range distance dependent anisotropies in the spin direction. These findings set constraints and show pathways for the design of $d^5$ materials that can host compass-like magnetic interactions.

\end{abstract}

\pacs{71.70.Ej, 71.70.Ch, 78.70.Ck, 78.70.En, 72.80.Ga}

\maketitle

\begin{figure}
\includegraphics[width=1.0\textwidth]{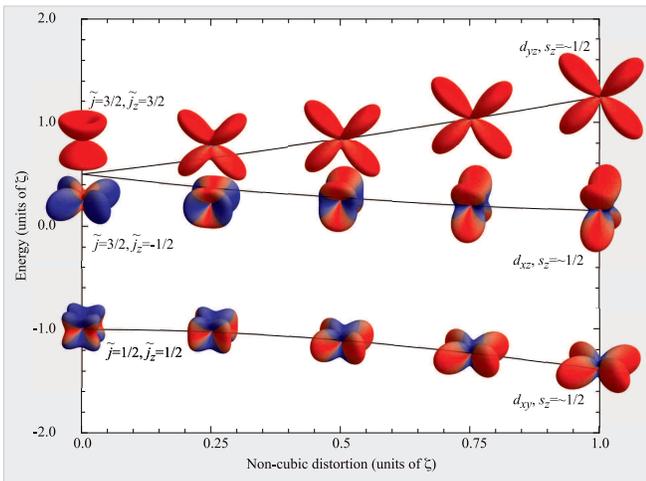}
\caption{(Color Online) Charge density plots of the local Ir-$t_{2g}$ orbitals including spin-orbit coupling as a function of non-cubic ($D_{2h}$) crystal-field splitting. Color refers to the spin-projection on the $z$ axis.}
\end{figure}

The class of 5$d$ transition metal oxides has attracted considerable attention in recent years. Exotic electronic states are expected to be found as a result of the presence of strong spin-orbit interaction in the $5d$ shell and associated entanglement of the spin and orbital degrees of freedom. In particular, one hopes to see signatures for compass exchange interactions in iridium $5d^5$ oxides, which then provides a route for novel frustrated systems and for the sought-after physical realization of the Kitaev model \cite{Jackeli:2009hz, Chaloupka:2010gi}, leading to a gapless spin liquid with emergent Majorana fermion excitations \cite{Kitaev:2006ik}. Furthermore, electronic ground states with non-trivial band topologies have been predicted, including topological Mott insulators \cite{Pesin:2010dg}, Weyl semimetals, or axion insulators \cite{Wan:2011hi, Hermanns:2015eb}.

In most materials the exchange interaction does not depend on the direction of the spin, i.e. $H \propto S_i\cdot S_j$ where the index $i$ and $j$ label different sites and $S$ represents the spin operator vector. The excitation spectra of $H$ has gapless modes related to the free rotation of all spins. Spin orbit coupling is usually a small perturbation adding single ion anisotropy and Dzyaloshinskii-Moriya interactions to $H$, thereby opening a gap in the excitation spectrum. For materials with locally cubic symmetry and strong spin-orbit coupling a unique situation arises \cite{Jackeli:2009hz, Chaloupka:2010gi}. The $t_{2g}$ orbitals are split by spin-orbit coupling into a $\widetilde{j}=1/2$ doublet and a $\widetilde{j}=3/2$ quartet. For a filling of 5 electrons the single hole in the $\widetilde{j}=1/2$ spin-orbital has a cubic charge density, with a strong entanglement between the spin and orbital degrees of freedom. The left bottom picture in Fig. 1 shows the charge density of a $\widetilde{j}=1/2$ orbital colored according to the spin projection in the $z$ direction. It is this strong entanglement between spin and orbital degrees of freedom that leads to pseudo spin quantization axes dependent directional hopping, i.e. the different components of the pseudo spin operator vector show different interactions for the different spatial directions.

A fundamental prerequisite for these theoretical predictions to be applicable is the realization of the many-body $\widetilde{J}=1/2$ state to properly describe the ground state of iridium oxides \cite{Kim:2008gi, Kim:2009tt}. It has been realized early on that non-cubic distortions will lead to a destruction of the spin-orbit entanglement and leave the system with conventional magnetic excitations \cite{Jackeli:2009hz}. Fig. 1 shows the evolution of the $t_{2g}$ spin-orbitals as a function of non-cubic distortions. A crystal-field strength ($\Delta_{t_{2g}}$) of the same order as the spin-orbit coupling strength ($\zeta$) will bring one to a situation where the orbitals are basically the real $t_{2g}$ orbitals.

To what extent  $\widetilde{J}=1/2$ state can be materialized is a matter of intensive investigations. The best estimate so far for the energy of the non-cubic crystal-field comes from resonant inelastic x-ray scattering (RIXS) experiments measuring directly the $d$-$d$ excitations between the different crystal-field levels. The estimates found for $\Delta_{t_{2g}}$ in Sr$_2$IrO$_4$ ($\Delta_{t_{2g}}$=-0.137 eV \cite{Kim:2014bt}), Ba$_2$IrO$_4$ ($\Delta_{t_{2g}}$=0.05 eV \cite{MorettiSala:2014fm}) and (Na,Li)$_2$IrO$_3$ ($\Delta_{t_{2g}}$= 0.1 eV \cite{Gretarsson:2013fp}) indicate that the tetragonal and trigonal crystal field energy splitting in these iridates is smaller than the spin-orbit coupling constant ($\zeta\sim$~0.4~eV). This in turn suggests that one should be in a situation where the compass-like magnetic models are realized. Similar results were obtained by quantum chemical calculations \cite{Bogdanov:2015de}.

Despite these findings, x-ray magnetic circular dichroism measurements were interpreted in terms of a strong mixing of the $\widetilde{j}=1/2$ and $3/2$ orbitals in Sr$_2$IrO$_4$ \cite{Haskel:2012ix}. Furthermore, pure Na$_2$IrO$_3$ and Li$_2$IrO$_3$ at ambient pressure were found to order magnetically in contradiction with the expected Kitaev spin-liquid state \cite{Singh:2010gg,Singh:2012db,Ye:2012dr,Modic:2014ch,Biffin:2014jz,Takayama:2015gm}. Various theoretical models, in particular the Kitaev-Heisenberg model, have been proposed, but a clear explanation of the magnetic ground state in the A$_2$IrO$_3$ iridates remain elusive \cite{Chaloupka:2010gi, Singh:2012db, Kim:2012dg, Chaloupka:2013ju, Yamaji:2014be}. Furthermore some theoretical studies challenged all together the picture of iridates having local moments, favoring a description in terms of molecular orbitals \cite{Mazin:2012ua, Foyevtsova:2013ks}.

In view of these puzzling and in-part contradicting reports in the literature, we set out to determine experimentally the local electronic structure of the iridium ion in Sr$_2$IrO$_4$, which is considered a model compound for the class of iridates. Here we use core-to-core RIXS as the experimental method of choice as we will explain in the next paragraph. We also aim to quantitatively explain the experimental spectra using one-electron parameters extracted from \textit{ab-initio} band structure calculations, with which we then can draw a more concise scheme of the low-energy Hamiltonian for the iridates.

\begin{figure}
\includegraphics[width=1.0\textwidth]{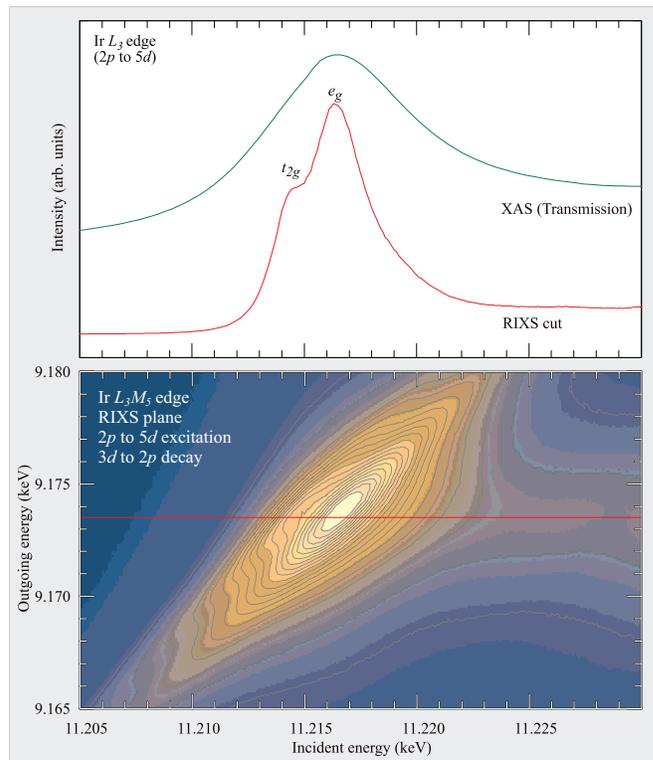}
\caption{(Color Online) Top panel: comparison of the x-ray absorption spectra, measured in transmission with RIXS data as a function of incident energy at constant outgoing energy. Bottom panel: intensity map of the RIXS intensity as a function of incident and outgoing energy.}
\end{figure}

One of the most direct methods to determine the local electronic structure of the transition metal ion in oxides is $L_{2,3} $ polarization dependent x-ray absorption spectroscopy: it is element specific and the dipole selection rules provide very selective and detailed information \cite{Tanaka:1994vx, deGroot:1994wi}. While successful for the study of $3d$ transition metal oxides, its use for the $5d$ systems is limited due to the fact that the lifetime of the $2p$ core hole is so short (due to the large amount of possible Auger decay channels of the $2p$ core hole for a $5d$ ion) that it washes out most of the fine multiplet details that make this type of spectroscopy so powerful. Using core-to-core $L_3M_5$ RIXS, we can circumvent this problem. The $2p_{3/2}$-to-$5d$ ($L_3$) excited state coherently decays via the $3d_{3/2}$-to-$2p$ ($M_5$) transition into an excited state of lower energy with a smaller amount of Auger decay possibilities and thus a much larger lifetime.

\begin{figure*}
\floatbox[{\capbeside\thisfloatsetup{capbesideposition={right,top},capbesidewidth=0.3\textwidth}}]{figure}[\FBwidth]
{\caption{(Color Online) Experimental and theoretical spectra. (r,g,b,y) description of the geometries used. (r,b) The incoming light is polarized in the \textit{ab} plane. (g,y) The incoming light is polarized parallel to \textit{c}. (r,g) The polarization of the outgoing light is perpendicular to the polarization of the incoming light. (b,y) The Poynting vector of the outgoing light is perpendicular to the polarization of the incoming light. (1-4) Experimental RIXS spectra, the fitted ionic model with unphysical values for the Slater integrals and the \textit{ab initio} calculation comparing different polarizations of the light. (1,2) Spectra taken with different polarization of the incoming beam. (3,4) Spectra taken with the same polarization of the incoming light, but with different polarizations of the outgoing light.}}
{\includegraphics[width=0.65\textwidth]{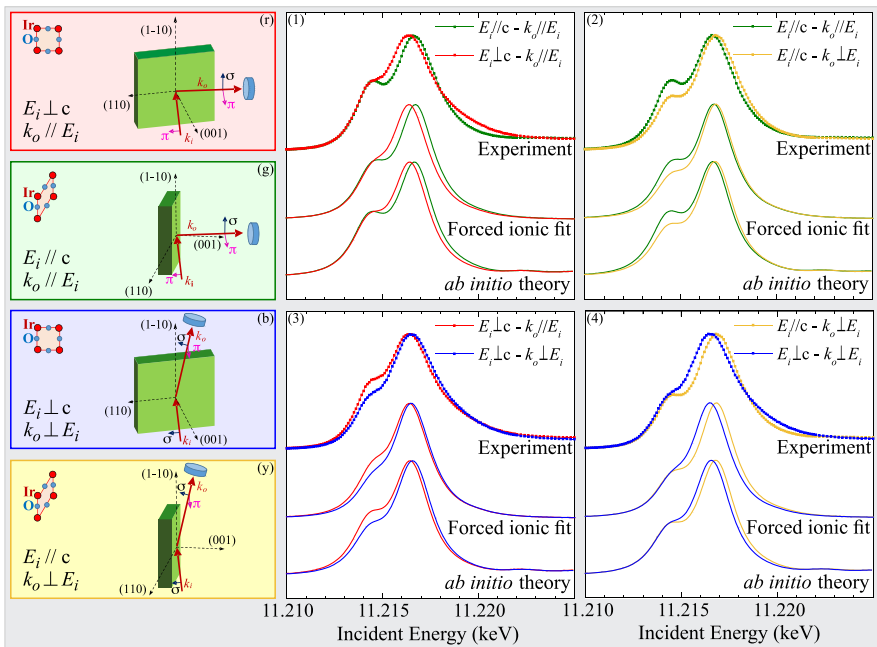}}
\end{figure*}

The results of the $L_3M_5$ RIXS on Sr$_2$IrO$_4$ are shown in the bottom panel of Fig. 2 as maps of the intensity as a function of the incident and outgoing photon energies \cite{supplementary}. The resonant excitation is indeed broad as a function of incoming energy, related to the above mentioned short $2p$ core hole lifetime. The spectra are, sharp in the energy loss, related to the much larger lifetime of the $3d$ core hole. We can take a cut through this map at constant emitted photon energy and retrieve spectra that are similar to x-ray absorption spectra, but with much sharper and better resolved features, as can be seen in the top panel of Fig. 2. These core-to-core $L_3M_5$ RIXS spectra show two peaks, which are roughly related to excitations into the $t_{2g}$ and $e_g$ orbitals.

In order to determine crystal field strengths and orbital occupations of the different orbitals we now vary the polarization of the incoming light \cite{deGroot:2008wo, Chen:1992ku, Haverkort:2004gg}. We hereby would like to note that RIXS spectral cuts are different from standard XAS \cite{deGroot:1994tz, vanVeenendaal:1996tb, Kurian:2012de} and that in particular the measured intensity depends on the polarization of the emitted photon. We utilize this and not only vary the polarization of the incoming light either in the $ab$ plane or in the $c$ direction, but we also measure the outgoing light with the detector either in the direction of the incoming polarization or perpendicular to it, thereby making use of the fact that light is always polarized perpendicular to the Poynting vector.

Fig. 3 shows the experimental data for the four different geometries we used. We show four panels (1-4), each one comparing two different spectra. If we change the incoming polarization from in-plane to out-of-plane we see that the $e_g$ derived peak shifts by about 0.3 eV, from which one can deduce that the energy of the $d_{z^2}$ orbital is higher than the energy of the $d_{x^2-y^2}$ orbital \cite{Haverkort:2004gg}. The shift of the $t_{2g}$ derived peak is small, confirming that the energies of the $t_{2g}$ orbitals are within a few 100 meV degenerate. Furthermore we observe that the peak positions and intensities change considerably depending on the outgoing polarization.

We now analyze the spectra quantitatively using an ionic crystal field model which contains the full Coulomb and exchange interaction between the Ir $\mathrm{5d}$, $2p$ and $3d$ electrons as well as the spin-orbit interaction and a non-cubic crystal-field. (Please note, in this letter, we will use Italic font for the atomic like Wannier orbitals and Roman font for the extended Wannier orbitals that describe the low energy eigen-states \cite{Haverkort:2012du, supplementary}.) The simulations from the ionic model are shown in panel (1-4) of Fig. 3 (labeled "Forced ionic fit"). We observe that the match between the experiment and simulations is excellent. As input parameters for the simulations, we placed the $d_{z^2}$ orbital about 0.5 eV higher in energy than the $d_{x^2-y^2}$ and the $d_{xz}$ / $d_{yz}$ only about 150 meV higher in energy than the $d_{xy}$. Yet, we must note that we had to reduce the Slater integrals, describing the Ir multipolar part of the Coulomb and exchange interactions, to 20\% of their atomic Hartree-Fock values. This is suspicious as one knows from an early spectroscopic study \cite{Antonides:1977wi} that these multipolar interactions are hardly ($\approx$80\%) reduced from their atomic Hartree-Fock values. The fact that one needs such a large reduction of the Slater integrals is a strong indication that the system is highly covalent.

\begin{figure*}
\floatbox[{\capbeside\thisfloatsetup{capbesideposition={right,top},capbesidewidth=0.3\textwidth}}]{figure}[\FBwidth]
{\caption{(Color Online) (a) Energy level diagram showing the interaction of the Ir $t_{2g}$ orbitals with the Ligand orbitals to form the bonding and anti-bonding $\mathrm{t}_{\mathrm{2g}}$ orbitals. (b) band-structure including the down-folded bands colored according to their Wannier function character. (c) Ir $d_{xz}$ and $d_{yz}$ partial density of states. (d) O $2p$ partial density of states of those orbitals that hybridize with the Ir $d_{xz}$ and $d_{yz}$ orbitals. (e) Ir $d_{xy}$ partial density of states. (f) O $2p$ partial density of states of those orbitals that hybridize with the Ir $d_{xy}$ orbital. (g,h) Charge density plots of the anti-bonding Ir $\mathrm{d}_{\mathrm{yz}}$ and Ir $\mathrm{d}_{\mathrm{xy}}$ orbital such that 90\% of the charge density is inside the contour. (k) Range dependence of the hopping for the Ir $\mathrm{d}_{\mathrm{xz}}$/$\mathrm{d}_{\mathrm{yz}}$ and $\mathrm{d}_{\mathrm{xy}}$ orbitals. The Ru-Ru directions are indicated in pseudo cubic notation.}}
{\includegraphics[width=0.65\textwidth]{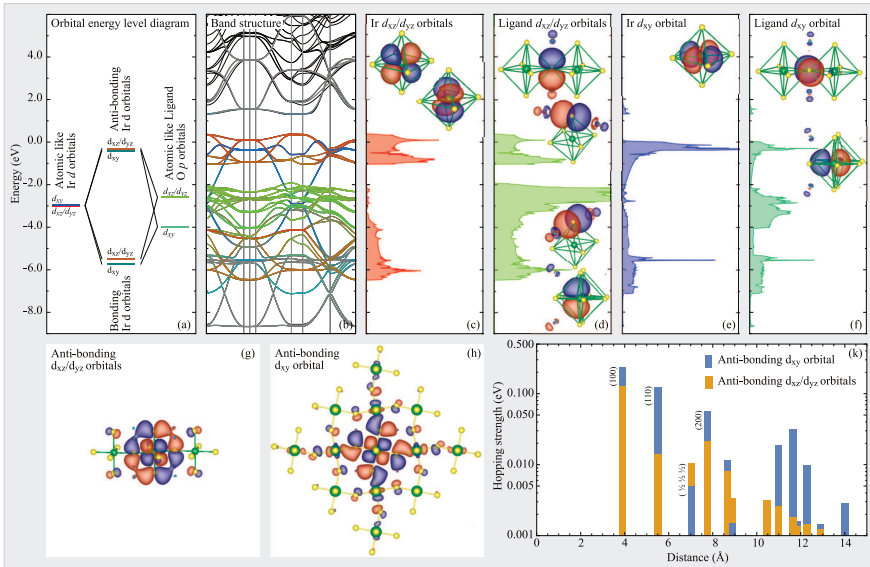}}
\end{figure*}

In the next step we analyze the spectra quantitatively using \textit{ab initio} multiplet ligand field theory (MLFT) calculations, where covalency is explicitly taken in account. In MLFT one creates a local cluster model including one correlated Ir $5d$ shell with dynamical charge fluctuations to ligand orbitals. We start our MLFT calculation by performing a density functional theory (DFT) calculation for the proper, infinite crystal using FPLO \cite{Koepernik:1999uw}. From the (self-consistent) DFT crystal potential we then calculate a set of Wannier functions suitable as single-particle basis. The Wannier basis and the DFT potential determine the one electron parameters of the local MLFT cluster model. Our procedure is similar to the method originally devised by Gunnarsson \textit{et al.} \cite{Gunnarsson:1990tw} and implemented in the code Quanty \cite{Haverkort:2012du, Haverkort:2016hz, supplementary}. The Ir 5$d$ Wannier orbitals are close to the atomic Ir DFT orbitals (Fig. 4(c,e) ). The ligand Wannier orbitals are similar to the atomic O $2p$ orbitals, but contain strong deformations due to the Ir $6s$ and Sr orbitals (Fig. 4(d,f) ). The ligand orbitals are not optimized to be localized or atomic like, but to capture the Ir $5d$ charge fluctuations as good as possible on a small as possible basis \cite{Haverkort:2012du, Lu:2014de}. Our method directly allows for the calculation of various forms of core level spectroscopy including RIXS \cite{Haverkort:2012du, Glawion:2011jf, Glawion:2012gs}. The calculated spectra are displayed in panel (1-4) of Fig. 3 (labeled "ab initio theory"). Similar to the ionic crystal field model, the MLFT calculations also show excellent agreement with the experiments, but in contrast to the former we now have used the full atomic Hartree-Fock values for the multipolar part of the Coulomb interaction. Thus, by including covalency in a realistic manner via MLFT, we can avoid the use of any \textit{ad-hoc} and unrealistic screening of the multipolar Coulomb interactions.

Having established that we can have an excellent simulation of the spectra using parameters that are determined from DFT \cite{parameters}, we can look at the implications of these calculations. Fig. 4(a) displays the DFT orbital energy level diagram. Our calculations reveal that the on-site energies of the It $t_{2g}$ Wannier orbitals are about the same, while the hopping of the $d_{xy}$ orbital to its ligand orbitals is larger than the hopping of the $d_{xz}$ and $d_{yz}$ orbitals to their ligand orbitals. This is in line with the tetragonal elongated local crystal structure. Nonetheless, the resulting energies of the anti-bonding Ir $t_{2g}$ - O $2p$ levels are practically degenerate, with the (Roman font \cite{Haverkort:2012du,supplementary}) $\mathrm{d}_{\mathrm{xz}}$ / $\mathrm{d}_{\mathrm{yz}}$ orbitals only about 150~meV higher in energy than the $\mathrm{d}_{\mathrm{xy}}$, thereby confirming the small difference found in the previous RIXS simulations. The reason that these large differences in the hopping do not result in a sizable energy splitting of the anti-bonding orbitals is the different on-site energy of the ligand O $2p$ orbitals (Fig. 4(a) ). We find a strong hopping to ligand orbitals at a slightly positive energy for the $d_{xy}$ and a weaker hopping to ligand orbitals at a slightly negative energy for the $d_{xz}$ and $d_{yz}$. The on-site energy of the ligands is thus crucial in the final determination of the anti-bonding state energies. In other words, the near degeneracy of the effective orbitals close to the chemical potential is the result of canceling interactions. A similar effect has been found in Mn doped Sr$_3$Ru$_2$O$_7$ \cite{Hossain:2008kb} and the importance of the non-local crystal structure has been pointed out for Sr$_2$IrO$_4$ and for pyrochlore iridates by Hozoi et al. \cite{Bogdanov:2015de,Hozoi:2014de}.

In setting up an appropriate low-energy Hamiltonian for the iridates, it is instructive to look at the spatial extent of the relevant Wannier functions. Fig. 4(g-h) plots the $\mathrm{d}_{\mathrm{xy}}$ and $\mathrm{d}_{\mathrm{xz}}$ / $\mathrm{d}_{\mathrm{yz}}$ effective orbitals.  One can clearly see that all orbitals are extending over the nearest-neighbour Ir sites and beyond. A similar statement can be made if one looks at the magnitude of the hopping interactions as a function of Ru-Ru distance as shown in Fig. 4(k). Needless to say that these effective orbitals are very different from the atomic orbitals shown in Fig. 1. The consequences are literally far reaching: the large extent of these effective orbitals leads to magnetic exchange couplings that are beyond nearest-neighbour only, adding extra terms that are not compatible with the Kitaev model. On top of this, the strong covalency amplifies the small on-site differences between the orbitals into strong orbital anisotropies for the effective low energy Hamiltonian: the $\mathrm{d}_{\mathrm{xy}}$ orbital is even more extended than the $\mathrm{d}_{\mathrm{xz}}$ and $\mathrm{d}_{\mathrm{yz}}$ due to the fact that the $\mathrm{d}_{\mathrm{xy}}$ can hybridize in two directions instead of one for the $\mathrm{d}_{\mathrm{xz}}$ and $\mathrm{d}_{\mathrm{yz}}$.

It thus becomes important to test the stability of compass models not only against the magnitude of the non-cubic distortions, but also to investigate the influence of anisotropic beyond-nearest-neighbor interactions. Alternatively, one can also look for iridate materials with crystal structures having large Ir-Ir interatomic distances so that the Ir-Ir exchange interactions can be well described by nearest-neighbor terms only.  A$_2$BB'O$_6$ double perovskites containing Ir$^{4+}$ and non-magnetic ions on the BB'-sites may form a good starting point. A challenge here will be to get the Ir and the non-magnetic ions to be highly ordered as to avoid spin-liquid phases due to disorder. Another option is to go to less charged $d^5$ ions, e.g. Ru$^{3+}$ or Os$^{3+}$ (OsPS, OsCl$_3$ or RuCl$_3$ for example) and/or to fluorine compounds to reduce the covalency. In these materials the charge transfer energy will be large and positive so that the materials are less covalent, thereby reducing the spatial extent of the anti-bonding $\mathrm{t}_{2g}$ Wannier orbitals and thus the range of magnetic interactions.

In conclusion, using core-to-core RIXS we find that the $\mathrm{t}_{2g}$ orbitals in Sr$_2$IrO$_4$ are indeed nearly degenerate. This degeneracy is not the result of a cubic local structure but is instead accidental due to the cancellation of a strong hopping to ligand orbitals at a small positive energy for the $d_{xy}$ and a weaker hopping to ligand orbitals at a small negative charge transfer energy for the $d_{xz}$ and $d_{yz}$. Important is that the spatial extent of the low-energy or effective $\mathrm{t}_{\mathrm{2g}}$ orbitals is large, thereby reaching the nearest neighbor Ir atoms and beyond, with the consequence that the effective low-energy Hamiltonian has long range terms, thereby also creating strong orbital anisotropies depending on the lattice structure.

\begin{acknowledgments}
We gratefully acknowledge the ESRF staff for providing beamtime. We thank Klaus Koepernik for useful discussions. K.-T. Ko acknowledges support from the Max Planck-POSTECH Center for Complex Phase Materials (No. KR2011-0031558). The research in Dresden was also supported by the Deutsche Forschungsgemeinschaft through SFB 1143.
\end{acknowledgments}

\end{document}